\documentclass[12pt,a4paper]{cibb}

\usepackage{subfigure,graphicx}
\usepackage{amsmath,amsfonts,latexsym,amssymb,euscript,xr}
\usepackage{booktabs}
\usepackage[nodayofweek]{datetime}
\usepackage{hyperref}

\usepackage[table]{xcolor}
\usepackage{color,colortbl,tabularx}

\usepackage[english]{babel}
\usepackage[protrusion=true,expansion=true]{microtype}
\usepackage{amsmath,amsfonts,amsthm}
\usepackage{pifont}

\definecolor{LightBlue}{rgb}{0.88,0.9,0.9}

\newcommand{\CellWithForcedBreak}[2][c]{
\begin{tabular}[#1]{@{}c@{}}#2\end{tabular}
}

\title{\Large $\ $\\ \bf \textit{mmid}: Multi-Modal Integration and Downstream analyses for healthcare analytics in Python}

\author{\large Andrea Mario Vergani$^{1,2,3}$, Valeria Iapaolo$^{3}$, Emanuele Di Angelantonio$^{1,4,5,6,7,8}$, Marco Masseroli$^{2}$, Francesca Ieva$^{1,3}$}
\address{\footnotesize $\ $\\$^1$ Human Technopole, Viale Rita Levi-Montalcini 1, 20157, Milan, Italy \\
$^2$ Department of Electronics, Information and Bioengineering (DEIB), Politecnico di Milano, Via Ponzio 34/5, 20133 Milan, Italy \\
$^3$ MOX lab, Department of Mathematics, Politecnico di Milano, Via Bonardi 9, 20133, Milan, Italy \\
$^4$ British Heart Foundation Cardiovascular Epidemiology Unit, Department of Public Health and Primary Care, University of Cambridge, Papworth Road, Cambridge Biomedical Campus, Cambridge CB2 0BB, UK \\
$^5$ Victor Phillip Dahdaleh Heart and Lung Research Institute, University of Cambridge, Papworth Road, Cambridge Biomedical Campus, Cambridge CB2 0BB, UK \\
$^6$ NIHR Blood and Transplant Research Unit in Donor Health and Behaviour, University of Cambridge, Cambridge, UK \\
$^7$ BHF Centre of Research Excellence, School of Clinical Medicine, University of Cambridge, Addenbrooke's Hospital, Cambridge, UK \\
$^8$ Health Data Research UK Cambridge, Wellcome Genome Campus and University of Cambridge, Cambridge, UK \\

\bigskip
}

\abstract{\small data integration, multi-modal health data, Python package. \normalsize
\\[17pt]
{\bf Abstract.} \textit{mmid} (Multi-Modal Integration and Downstream analyses for healthcare analytics) is a Python package that offers multi-modal fusion and imputation, classification, time-to-event prediction and clustering functionalities under a single interface, filling the gap of sequential data integration and downstream analyses for healthcare applications in a structured and flexible environment. \textit{mmid} wraps in a unique package several algorithms for multi-modal decomposition, prediction and clustering, which can be combined smoothly with a single command and proper configuration files, thus facilitating reproducibility and transferability of studies involving heterogeneous health data sources. 
A showcase on personalised cardiovascular risk prediction is used to highlight the relevance of a composite pipeline enabling proper treatment and analysis of complex multi-modal data. 
We thus employed \textit{mmid} in an example real application scenario involving cardiac magnetic resonance imaging, electrocardiogram, and polygenic risk scores data from the UK Biobank. 
We proved that the three modalities captured joint and individual information that was used to (1) early identify cardiovascular disease before clinical manifestations with cardiological relevance, and (2) do it better than single data sources alone. Moreover, \textit{mmid} allowed to impute partially observable data modalities without considerable performance losses in downstream disease prediction, thus proving its relevance for real-world health analytics applications (which are often characterised by the presence of missing data).
}

\begin{document}
\thispagestyle{myheadings}
\pagestyle{myheadings}

\bigskip
\bigskip
\noindent
\textbf{Corresponding author:} Andrea Mario Vergani \href{mailto:andreamario.vergani@polimi.it}{[andreamario.vergani@polimi.it]}
\\
\textbf{Code repository:} \url{https://github.com/ht-diva/mmid}
\bigskip
\bigskip

\section{\bf Introduction}
\label{sec:SCIENTIFIC-BACKGROUND}
The increasing availability of 
multi-modal data is revolutionising healthcare~\cite{hao2025,tabakhi2023}, offering invaluable opportunities to study the impacts of 
heterogeneous health data types on biological mechanisms and disease onset
. In this context, data integration techniques have been attracting growing interest, with different multi-modal data fusion approaches proposed in the literature. These methodologies aim to bridge the gap between data collection from various sources and the need to shift from a single- to a multi-modal perspective for precision health analytics~\cite{tabakhi2023}, and were successfully applied especially in the integration of omics on the DNA-to-protein pathway~\cite{Argelaguet2018,jive,FENG2018241,10.3389/fonc.2020.01065,milite}. Despite the presence of several frameworks for data integration in healthcare 
(e.g., based on matrix factorisation, deep representation learning, \ldots), 
algorithm selection is generally not trivial for various reasons: (1) data fusion is mainly unsupervised, so its result cannot be compared with a ground truth and is intrinsically neither right nor wrong; (2) data integration is usually an intermediate analysis step preceding the downstream task of interest (e.g., patient clustering, disease prediction), thus its evaluation is not independent of the latter; 
(3) due to the previous points, the evaluation of data fusion is complex and requires structured comparisons of methods, in terms of downstream performance, biological meaning, and interpretability of the results; 
(4) data integration algorithms have not been extensively tested with complex health data sources (especially non-omics ones, e.g., medical imaging or biomedical signals) in the literature.\\
To address these challenges, we propose \textit{mmid} (Multi-Modal Integration and Downstream analyses for healthcare analytics), a 
Python package 
that combines multiple integration 
and downstream 
algorithms 
in a single framework, thus allowing easier application, comparisons, and evaluations of data fusion paradigms in different scenarios. The 
package 
originates mainly as a response to three unmet needs in the multi-modal integration field: (1) the impossibility to access state-of-the-art multi-modal fusion algorithms in a unified way (indeed, every integration method generally comes with its own specific library/interface)~\cite{Iapaolo}; (2) the lack of tools to guarantee the transferability of the studies, as well as the critical assessment of the methods employed and their easy comparison with alternative algorithms; (3) the difficulty to evaluate the marginal contribution of single modalities to a downstream task, in a multi-modal context. To address these, \textit{mmid} collects some of the main state-of-the-art multi-modal integration methods, and connects their outputs directly to the subsequent tasks of interest, under a unique 
Python package. This allows users to test and compare different algorithms and downstream analyses from a single standardised interface, thus enhancing the transferability of results and the assessment of the impacts of methods and data sources across tasks. 
Furthermore, \textit{mmid} facilitates data handling and input formatting, which is otherwise specific to the package of the single selected integration algorithm, allowing the user to test different fusion methods on the same tabular datasets (i.e., without any algorithm-specific input modifications needed); indeed, input formatting happens transparently within the \textit{mmid} package. 
Finally, \textit{mmid} is smoothly extensible with further integration approaches and downstream algorithms and tasks, thus enabling it to easily stay up-to-date with new advancements in the literature and state-of-the-art methodologies; as an example, adding a new integration method only requires to properly wrap it into the $Integrator$ class of \textit{mmid}; analogously, further downstream models can be inserted simply extending the $PredictionModel$ class of the package (specifically, its $ClassificationModel$, $SurvivalModel$ and $ClusteringModel$ subclasses for classification, survival and clustering algorithms, respectively).\\
Compared to similar Python or R packages already available in the literature, \textit{mmid} directly connects data fusion and downstream analyses for healthcare applications, it is currently up-to-date and extensible to the latest developed algorithms in the multi-modal fusion field, and its source code is freely available for anyone to use or build upon (see\textbf{~\hyperref[sec:AVAILABILITY]{Code availability}}). 
Popular existing software for multi-modal fusion provides a range of algorithms for integration without connecting them to downstream tasks, thus making their comparison on real case studies more elaborate and potentially dependent on algorithm-specific code, cohort management, \ldots 
For example, \textit{mvlearn}~\cite{perry2021mvlearn} and \textit{muon}~\cite{muon} - two Python frameworks for multi-modal data management and fusion - are restricted to integration only (i.e., they do not solve the problem of structured downstream analyses). Similarly, \textit{mixOmics}~\cite{10.1371/journal.pcbi.1005752} is an R package for fusion only (again, no downstream analyses included), which is also not up-to-date with the latest developed algorithms in the field. The \textit{AstraZeneca - artificial intelligence (AZ-AI) multimodal pipeline}~\cite{nikolaou} in Python actually bridges between multi-modal feature integration and downstream survival prediction, but it is not publicly available as a package or code repository, and its integration algorithms are specific to cancer applications. In contrast, \textit{mmid} solves the issue of connecting fusion and general downstream tasks for healthcare (including clustering, time-to-event and disease classification predictions) in a public and easy-to-use Python package, enabling structured analyses through a single function call.\\
We tested \textit{mmid} in a real case study 
for the prediction of 
cardiovascular risk using cardiac imaging, electrocardiogram (ECG), and genetic data from the UK Biobank (UKB). 
In this 
example application study, we aimed to show that integrating such 
heterogeneous 
modalities can be profitable for personalised cardiovascular risk assessment, and can generally pave the way for a better informed combination of omics and medical test data in a variety of clinical and biological tasks. 
We focused on the cardiovascular risk prediction domain because it is intrinsically highly multi-modal, and an extensive assessment of the marginal contribution of various cardiological tests across disease subtypes is currently lacking. 
Specifically, while in previous studies
~\cite{Mayala_2018,si2024,10.1371/journal.pmed.1003498} cardiac imaging, ECG and genetics were shown to have a clear role in the 
definition 
of cardiovascular diseases (CVDs) when considered alone, their joint impact is still unclear, especially with respect to the risk for healthy subjects to develop disease in the future. 
More in general, our 
application study exemplifies a domain in which pipelines to guarantee standardisation of analytical approaches and reproducibility of results were lacking; in such a situation, \textit{mmid} facilitated our analyses, allowing easy extensions from single experiments to structured comparison of algorithms, tasks and results across endpoints.\\
The rest of the paper is structured as follows:\textbf{~\autoref{sec:pipeline}} describes the \textit{mmid} Python package, its modules and its main function to run the analyses; in\textbf{~\autoref{sec:case-study}} we present an example application study, detailing the modalities employed in the analyses and the CVD endpoints, the sequential steps of our integration and disease prediction framework, and the selected algorithms within \textit{mmid}; in\textbf{~\autoref{sec:RESULTS}} we present the results of the example case study, showing that \textit{mmid} helps disentangling the joint and individual contributions of modalities to the risk across disease subtypes, and that it is possible to provide predictions even for subjects with partially collected data sources; finally, in\textbf{~\autoref{sec:discussion}} we discuss the impacts, strengths and limitations of our work, before concluding with some final remarks in\textbf{~\autoref{sec:CONCLUSIONS}}.\\

\section{\bf The \textit{mmid} package}
\label{sec:pipeline}
\textit{mmid} is a 
Python package 
coded in Python $3.7.12$, which allows the combination of multiple healthcare data sources and the execution of a downstream task in a selected cohort based on the merged representation of the multi-modal input datasets. 
We present the structure of \textit{mmid} in\textbf{~\autoref{sec:structure}}, while the main function to run the analyses is described in\textbf{~\autoref{sec:mmid_function}}.

\subsection{\bf \it \textit{mmid} structure}
\label{sec:structure}
\begin{figure}[h]
\vspace{3mm}
 \begin{center}
 \includegraphics[width=\textwidth]{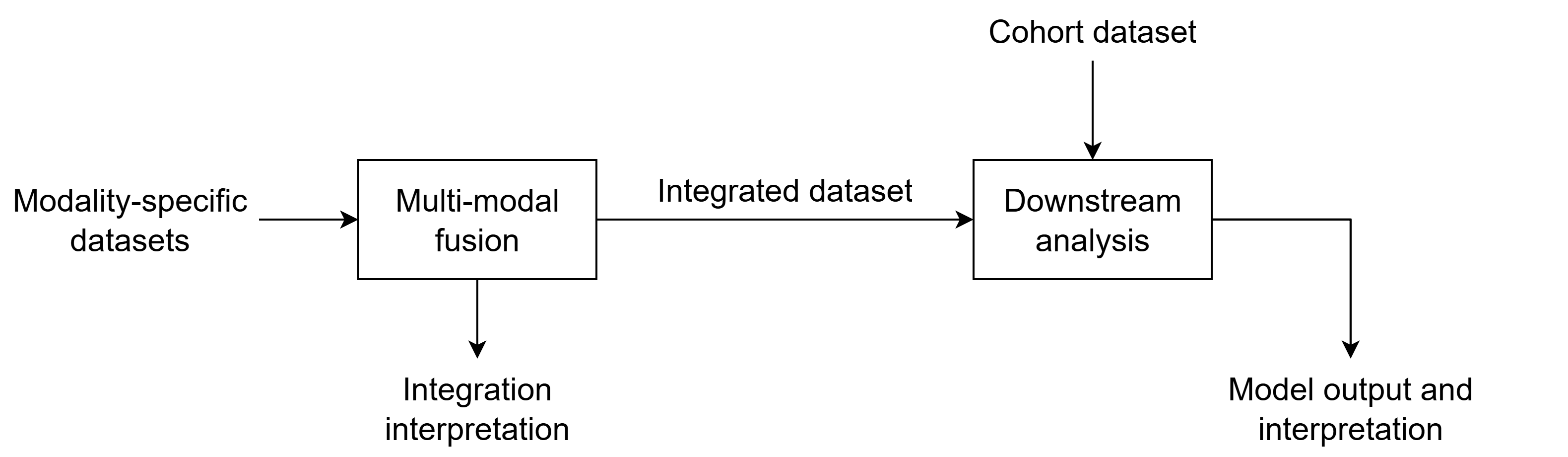}
\caption{\textbf{\textit{mmid} outline.} \textit{mmid} flow, with sequential multi-modal integration and downstream analysis, and their inputs and outputs.
\label{fig:outline}}
 \end{center}
\end{figure}

The \textit{mmid} package is composed of the following sequential modules, as depicted in\textbf{~\autoref{fig:outline}}:
\begin{enumerate}
    \item The \textbf{\textit{Multi-modal fusion}} module (corresponding to the $Integrator$ Python class), which takes as input some modality-specific tabular datasets and combines them into a merged representation in an unsupervised way.
    \item The \textbf{\textit{Downstream analysis}} module (corresponding to the $PredictionModel$ Python class), which uses the merged representation to fit a disease classification, time-to-event prediction, or patient clustering analysis on a cohort passed as input. The cohort dataset should include disease information for a set of subjects completely or partially overlapping the set of individuals in the integrated dataset.
\end{enumerate}
The key strengths of 
\textit{mmid} 
reside in the fact that: (1) it is 
general and 
agnostic to the disease domain, allowing the integration of any number and kind of tabular modality-specific datasets (e.g., omics, imaging-derived features, \ldots); (2) it is configurable, meaning that the user can easily select the type of downstream task and the algorithms for data fusion and disease classification / time-to-event prediction / clustering through proper configuration files (see\textbf{~\autoref{sec:mmid_function}}). 
The main strict input requirement for \textit{mmid} is the availability of modality-specific datasets in tabular forms (which may somehow need to be harmonised, if necessary, before being passed as inputs to the package).

\subsection*{\bf \it Multi-modal fusion: the $Integrator$ class}
\label{sec:fusion}
The \textit{Multi-modal fusion} module of \textit{mmid} - embodied by the $Integrator$ class in Python - acts as a wrapper for unsupervised multi-modal fusion approaches that project whichever number of modality-specific 
tabular 
datasets into a single merged representation, which in turn becomes the input of the \textit{Downstream analysis} module. The user can decide to integrate the input datasets through 
different approaches, ranging from simple joining up to more sophisticated methodologies. According to the types of data and purposes of the downstream analyses, the following options are available: 
(1) an early fusion approach; (2) the Multi-Omics Factor Analysis (MOFA+) algorithm~\cite{Argelaguet2020}; (3) the Angle-based Joint and Individual Variation Explained (AJIVE) method~\cite{FENG2018241}. Specifically:
\begin{enumerate}
    \item \textbf{Early fusion} directly concatenates the input datasets, so the resulting merged representation is simply the concatenation of all the original modality-specific features~\cite{tabakhi2023}. 
    With the \textit{mmid} 
    package, the user can perform early fusion in a purely concatenation-based way, or preceded by a Principal Component Analysis (PCA) step on each modality independently to reduce the dimensionality of the resulting merged representation. Early fusion is an extremely simple and fast integration strategy, which however lacks in effectively identifying complex interactions; it is thus suited as a trivial strategy to combine a low number of features coming from (very) few modalities.
    \item \textbf{MOFA+} is a multi-modal integration algorithm that factorises $n$ heterogeneous datasets $Y_{i}$ ($N$ samples $\times$ $D_{i}$ features) into the product of a cross-modal matrix $Z$ ($N \times M$ - see below) and modality-specific weight matrices $W_{i}$ ($D_{i} \times M$), plus residual $\epsilon_{i}$, according to the following formulation:
    \begin{equation}
        Y_{i} = Z W^\top_{i} + \epsilon_{i}
    \end{equation}
    $(i=1, \dots, n)$, in which $Z$ represents the fusion factor matrix $N \times M$, with $M$ being the dimensionality of the integrated space corresponding to the principal sources of cross-modal variability. Thus, MOFA+ projects the datasets $Y_{i}$ into a $M$-dimensional 
    space, whose factors can be interpreted through the weight matrices $W_{i}$. Two possible ways to set $M$ are: (1) choose a value a priori, in which case a good practice is to determine how many $m_i$ principal components (PCs) explain $P\%$ of modality-specific variance (choosing $P$ with the same heuristics used in PCA), and define $M = \sum_i m_i$; (2) set $M$ to a sufficiently high value, then remove cross-modal dimensions explaining $< p\%$ of variance (with $p$ typically between $1\%$ and $5\%$). The choice of MOFA+ as fusion algorithm is particularly suited in case of partially observable data sources: indeed, the method is able to build $Z$ even in case of missing modalities for some of the analysed subjects. The implementation of MOFA+ in 
    \textit{mmid} 
    relies on the \textit{mofapy2} (version 0.6.7) and \textit{mofax} (version 0.3.7) Python packages~\cite{Argelaguet2020}.
    \item \textbf{AJIVE} is an extension of Joint and Individual Variation Explained (JIVE)~\cite{jive}, a popular multi-modal fusion method based on the following decomposition of the $n$ original datasets $Y_{i}$ $(i=1, \dots, n)$:
    \begin{equation}
        Y_{i} = J_{i} + I_{i} + R_{i} = S U^\top_{i} + S_{i} V^\top_{i} + R_{i}
    \end{equation}
    Basically, a JIVE-based 
    representation is composed of $M$ joint components (indicated with $J$ in the following), whose score matrix $S$ ($N \times M$) is common to all modalities (with modality-specific weights contained in the $D_{i} \times M$ matrices $U_{i})$, and $\sum_{i=1}^{n} M_{i}$ (i.e., $M_{1}$ components related to the first modality, \dots, $M_{n}$ components to the $n$-th modality) modality-specific individual components (indicated with $I$ in the following), defined by the corresponding weight matrices $V_{i}$ ($D_{i} \times M_{i}$) and score matrices $S_{i}$ ($N \times M_{i}$)~\cite{jive}. Thus, the method integrates the datasets $Y_{i}$ into a manifold of dimensionality $M + \sum_{i=1}^{n} M_{i}$, whose components can be clearly attributed to the joint effects explaining patterns across multiple modalities, or to the individual effects related to modality-specific variability. Furthermore, the AJIVE solution requires orthogonality between individual matrices and the joint space, ensuring that the joint and individual patterns captured by the method are unrelated~\cite{jive,FENG2018241}. While the values of $M$ and $M_i$ are determined during training, the user can decide how many modality-specific components $m_i$ to retain in the dimensionality reduction step preceding the projection to the cross-modal representation~\cite{FENG2018241}; to set good values of $m_i$, the best practice is to refer to the selection methods for the number of PCs in PCA (e.g., elbow method). Contrary to MOFA+, AJIVE allows to clearly decouple the contributions of the data sources to the cross-modal representation, thus explicitly identifying joint and modality-specific effects and interpreting them in terms of the original measured features. The AJIVE implementation included in 
    \textit{mmid} 
    is based on the \textit{mvlearn} Python library~\cite{mvlearn} (version 0.5.0).
\end{enumerate}
We included concatenation-based early fusion for all those cases where decoupling and/or more complex fusions of modalities are not required or desired, 
so it can be used as a simple benchmark against other integration methods. Furthermore, among other multi-modal fusion frameworks, we decided to wrap MOFA+ and AJIVE in \textit{mmid} because they are completely interpretable and they integrate datasets independently of downstream cohort and task. Specifically about interpretability, we considered it a key factor for healthcare applications, to explain the impact of the cross-modal representation in downstream analyses and how the merged features relate to the original data modalities. Indeed, the \textit{Multi-modal fusion} module always outputs the explanations behind the constructed merged representation (i.e., MOFA+ and AJIVE weight matrices), together with the latter itself (\textbf{\autoref{fig:outline}}). Last but not least, both MOFA+ and AJIVE have already been successfully employed in the healthcare domain~\cite{Argelaguet2018,jive,FENG2018241,Argelaguet2020}.\\
Eventually, while the merged representation is generally built on the 
samples for which all the data modalities have been collected, in the case of MOFA+ 
our package offers the possibility to impute in the cross-modal space those views that are missing for some of the observed subjects (by passing $latent\_impute = True$ to the $mmid()$ function of the package - see\textbf{~\autoref{sec:mmid_function}}). This enables the extension of the downstream analyses to larger cohorts, including individuals with availability of only portions of the full spectrum of data sources considered. We believe that such an option perfectly suits real-world applications and scenarios, in which subjects usually undergo specific medical tests depending on their health conditions. 
Note that integrating even in the case of partially observed modalities within \textit{mmid} is only available for MOFA+, 
since the other algorithms were not designed to deal with missingness.

\subsection*{\bf \it Downstream analysis: the $PredictionModel$ class}
\label{sec:downstream}
The merged representation of multi-modal health data built by the \textit{Multi-modal fusion} module enters the \textit{Downstream analysis} module of 
\textit{mmid}, where it is used to fit a downstream task on a selected cohort. The Python class representing this module of the pipeline is named $PredictionModel$. The downstream tasks implemented in 
\textit{mmid} 
are:
\begin{itemize}
    \item \textbf{Binary classification} ($ClassificationModel$ subclass), aimed at classifying the 
    occurrence of disease in the next $years\_risk\_classification$ years based on the merged representation of multi-modal health data, for the subjects of the cohort (comprising positive cases and negative controls) healthy at a given baseline. Although the basic prediction relies exclusively on the features of the cross-modal representation, the user can decide to include additional covariates from the cohort dataset in the model 
    (through the $cohort\_cov$ input parameter of the main function call - see\textbf{~\autoref{sec:mmid_function}}). Moreover, the user can configure the prediction time horizon by setting the value of $years\_risk\_classification$ 
    (as argument of the $mmid()$ function of the package - see\textbf{~\autoref{sec:mmid_function}}). 
    The classification models the user can employ include logistic regression from the \textit{statsmodels} module~\cite{statsmodels} (version 0.13.5), XGBoost from the \textit{xgboost} Python library~\cite{xgboost} (version 1.6.2), and some other algorithms (e.g., Naive Bayes, AdaBoost) from \textit{scikit-learn}~\cite{scikit-learn} (version 1.0.2).
    \item \textbf{Time-to-event prediction} ($SurvivalModel$ subclass), in which the merged representation is used as input of a survival analysis aimed at predicting the time to the first occurrence of a disease 
    for the subcohort of subjects healthy at baseline (in order to study incident events). 
    Also in this task the user can add 
    further covariates 
    and select among various survival algorithms, including the Cox proportional hazards (Cox-PH) model (from the \textit{lifelines} Python library~\cite{Davidson-Pilon2019} version 0.27.8), Random Survival Forest (from \textit{scikit-survival}~\cite{sksurv} version 0.17.2), XGBoost (from \textit{xgboost}~\cite{xgboost} version 1.6.2), DeepSurv~\cite{Katzman2018} (from \textit{pysurvival}~\cite{pysurvival_cite} version 0.1.2), and DeepHit~\cite{lee} (from \textit{pycox}~\cite{JMLR:v20:18-424} version 0.2.3).
    \item \textbf{Patient clustering} ($ClusteringModel$ subclass), where the subjects in the cohort are clustered according to their representation in the merged space of data modalities. Featured clustering algorithms in 
    \textit{mmid} 
    include DBSCAN and K-means from the \textit{scikit-learn} Python library~\cite{scikit-learn} (version 1.0.2).
\end{itemize}
Basically, the user can select the downstream task to perform, the algorithm, and the value of 
non-learnable 
hyperparameters through proper configuration files (see\textbf{~\autoref{sec:mmid_function}}). The idea is that several combinations of 
fusion algorithms, 
tasks and models can be tested just by changing the parameters and inputs to the 
package's main function, in order to understand the impacts of different modalities on health conditions and diseases when integrated together. 
Finally, the \textit{Downstream analysis} module outputs the interpretation of the model 
(\textbf{\autoref{fig:outline}}), based on feature weights and confidence intervals (e.g., in the case of logistic regression or Cox-PH model), or feature importance (e.g., for random forest, \ldots).

\subsection{\bf \it The $mmid()$ function to run the analyses}
\label{sec:mmid_function}
To run a multi-modal integration and downstream analysis with the \textit{mmid} package, the user can simply run the $mmid()$ function, which takes the following obligatory arguments:
\begin{itemize}
    \item $config\_data$: the path of the YAML data configuration file, in which the user specifies all the details on the input modality datasets (e.g., file paths, features to consider, number of factors $m_i$ in the integration).
    \item $config\_model$: the path of the YAML model configuration file, where the user selects the integration and downstream algorithms (and their hyperparameters) for analysis. For instance, to run multi-modal fusion with AJIVE followed by penalised logistic regression for disease classification, the user can create a YAML model configuration file specifying $use{:}$ $True$ under the $integration{:}$ $ajive{:}$ and $prediction{:}$ $logregrssm{:}$ keys, and indicate the L1 regularisation weight of $1.0$ (as an example) with $alpha{:}$ $1.0$ under the $prediction{:}$ $logregrssm{:}$ $params{:}$ key. The user then passes the path of this YAML file to $config\_model$ to run the analysis.
    \item $cohort\_path$ and $cohort\_file$, which identify the input cohort file.
    \item $end\_study\_date$ to declare the last observation date (i.e., end of the study).
    \item $out\_path$: the path where the results generated by \textit{mmid} are stored.
\end{itemize}
When called, the $mmid()$ function processes the input modalities passed in the data configuration file, integrates them through the \textit{Multi-modal fusion} module according to the method specified in the model configuration file, and runs the downstream classification/survival/clustering prediction (declared in the model configuration file) on the input cohort thanks to the \textit{Downstream analysis} module. The full results of the function call and analysis (comprising integration interpretation and downstream model output) are stored in the location passed through the $out\_path$ argument.\\
Additionally, the $mmid()$ function accepts optional arguments that allow the user to consider further covariates for the downstream analysis ($cohort\_cov$ argument), solve missingness in the latent space if possible ($latent\_impute$ argument), set the test size and the number of folds for k-fold cross-validation ($test\_size$ and $n\_folds$ arguments), \ldots\\
By properly setting configuration files and argument values to the function, the user can perform targeted analyses starting from the same (or different) sets of data and compare the results. For instance, changing data configuration while maintaining model configuration enables the testing of the same integration and downstream combination of algorithms on different sets of input data (e.g., single versus multiple modalities together); similarly, modifying model configuration while keeping the same data configuration allows the user to compare different integration and/or downstream methods on the same data (e.g., classification versus survival predictions, or AJIVE versus MOFA+, or logistic regression versus Naive Bayes). This facilitates the reproducibility of the studies and the structured evaluation of methodologies and experiments in the multi-modal integration setting for healthcare applications.

\section{\bf Application: Materials and methods}
\label{sec:case-study}

We exemplified the relevance of the \textit{mmid} 
package 
and, more generally, of the construction of a merged representation of complex health data for disease prediction through a significant and representative application study about the fusion of cardiac imaging, ECG and genetic data for the prediction of the future occurrence of CVD in healthy subjects.\\
In the cardiovascular field, large-scale complex data is increasingly collected 
and available through registries, biobanks and electronic health records. 
However, the interplay between 
e.g., genetic samples, imaging, signals and clinical records, and the way they interact to define CVD risk are still poorly explored.
~\cite{10.1093/eurheartj/ehaf947} recently combined clinical, metabolomic and polygenic scores for CVD risk prediction using feature selection and linear predictors, without exploring multi-modal fusion methodologies to manage heterogeneous inputs. To the best of our knowledge, despite the success of fusion methods to aggregate molecular and omics data, multi-modal integration has never been employed in the literature to combine the results of different medical examinations toward a better understanding of disease prognosis.\\
In our case study, we focused on CVDs. 
Their occurrence is influenced by a variety of factors (e.g., family history, smoking habits)~\cite{framingham,Hippisley-Coxj2099,fh}, so different modalities are expected to provide complementing perspectives on the cardiovascular health status of a subject. We thus employed the \textit{mmid} 
package to integrate cardiac magnetic resonance (CMR) imaging, ECG and genetic-derived features and use the corresponding representation to predict the future occurrence of CVD in healthy populations. With this 
application study, we aim to show the relevance of data integration in the task, as well as to disentangle the joint and individual contributions of modalities to CVD risk definition relying on \textit{mmid}.\\
In this section, we fully describe the data, the preprocessing steps, and how 
\textit{mmid} 
was used for our case study.

\subsection{\bf \it Data and preprocessing}
\label{sec:data}
We analysed data from the UKB, a United Kingdom-based biobank study involving about 500,000 subjects aged 40 to 69 at recruitment (2006–2010), who have been followed up to date linking general practitioner, death, hospital and cancer national registries. Furthermore, subsets of UKB participants underwent genetic testing, medical imaging, ECG, and other (repeated) health 
examinations 
throughout the observation period~\cite{ukb.10.1371/journal.pmed.1001779}.\\
For our multi-modal integration analyses, we considered CMR phenotypes, ECG-derived features, and genetic predisposition data available in UKB (\textbf{\autoref{fig:mod_overlap}}). In particular, we analysed: (1) the dataset of $84$ cardiac imaging measures extracted by~\cite{bai2020} on 31,923 subjects from UKB long and short axis CMR dynamic ECG-synchronised acquisitions; (2) the dataset of $12$ ECG measures (i.e., ventricular rate, P duration, PP interval, PQ interval, number of QRS complexes, QRS duration, QT interval, QTC interval, RR interval, P axis, R axis, T axis) directly derived during the ECG 
examination, available for 43,903 UKB participants; (3) the dataset of $36$ polygenic risk scores (PRSs) computed by~\cite{prs.10.1371/journal.pone.0307270} from genetic data of 485,906 UKB subjects, where a PRS represents the genetic predisposition of a person for a trait. Available traits for which PRSs were calculated included both continuous phenotypes (e.g., height) and diseases (e.g., atrial fibrillation). All three datasets were complete: if a modality was available for a participant, all the features for that modality were collected and reported (i.e., absence of intra-modality missing values). While the CMR and ECG datasets were intrinsically cardiovascular, the PRS one was indirectly linked to CVD, with about one third of available PRSs related to a cardiometabolic risk factor (e.g., hypertension) or condition (e.g., coronary artery disease); furthermore, CMR and ECG measures are directly measurable phenotypes, while PRSs reflect aggregate genetic predispositions. The CMR and ECG datasets were acquired during the same visit - the first UKB imaging visit (i.e., same subject-specific time point), while PRSs are fixed at conception.\\
We independently preprocessed the three modality-specific datasets by iteratively removing collinear features with a variance inflation factor $>10$, as collinearity is not recommended for data integration and prediction. We thus retained $68$ CMR-derived phenotypes, $9$ ECG measures, and all the $36$ PRSs (which were already lowly correlated). This feature preprocessing step is not part of \textit{mmid}, but was applied ad-hoc to our application study.\\
We then defined and dated the occurrence of four CVD subtypes - atrial arrhythmia (AA), coronary artery disease (CAD), general CVD and structural heart disease (SHD), i.e., the targets of our disease prediction tasks - based on hospital diagnoses, operative procedures, and causes of death available in UKB, following\textbf{~\autoref{tab:endpoint}}.

\subsection{\bf \it Data integration with mmid}
\label{sec:integration}
We used MOFA+ and AJIVE 
within \textit{mmid} 
to integrate the PRS, ECG, and CMR measures datasets into a low-dimensional representation capturing the main sources of variability across the three modalities. 
When employing AJIVE, we decided to retain a number of components explaining at least $P=80\%$ of modality-specific variance in the initial dimensionality reduction step (i.e., the step preceding the projection to the joint and individual components of the cross-modal representation, as described by~\cite{FENG2018241}): this corresponded to $m_1=25$, $m_2=6$ and $m_3=24$ components for the CMR, ECG, and PRS datasets, respectively. In the case of MOFA+, instead, we projected the modalities into a 10-dimensional merged space, and discarded the cross-modal features that explained $< p = 5\%$ of variance (for each modality independently). The selected cutoffs for explained variance follow from exploratory analyses at the basis of our previous work on UKB clinical, ECG and cardiac imaging data~\cite{Iapaolo}, 
with the idea for this application study to reduce the dimensionality maximally within reasonable threshold values. In general, the optimal cutoffs are problem-specific and should be fine-tuned depending on the data and methods used.\\
To study the impacts of modalities on prediction across 
CVD subtypes, we primarily focused 
on 
the AJIVE integration, 
which 
explicitly disentangled the joint contribution related to the interplay of modalities and modality-specific components, thus directly clarifying the role of data sources in the integration and downstream tasks. We then replicated the analyses employing MOFA+ to investigate the effects of missing data sources (for some subjects) and the corresponding impacts on disease prediction performance, leveraging the modality imputation feature offered by the method.\\
In both cases, we inspected the integration weight matrices 
generated by \textit{mmid} 
to interpret the interplay of the modalities defining the cross-modal representation. 
This representation was then used - always within our package and in smooth continuity with the integration step - 
to fit our disease prediction analyses.

\subsection{\bf \it Disease prediction with mmid}
\label{sec:prediction}
We focused on the prediction of the future occurrence of CVD subtypes in healthy subjects under two different scenarios, i.e., (1) incident disease classification in a fixed time horizon of $years\_risk\_classification = 5$ years - evaluating performance through the area under the receiver operating characteristic curve (AUC), and (2) time-to-event prediction in a survival analysis framework - relying on the concordance index (in the implementation of the \textit{lifelines} Python library version 0.27.8) for model evaluation. The survival cohort sizes were bigger than those for the classification scenarios, mainly because not all subjects were observed for at least $years\_risk\_classification = 5$ years (in that case, \textit{mmid} automatically excluded them from the disease classification analyses). In both cases, we 
entirely relied on our package to 
assess 
the value of AJIVE multi-modal integration for the tasks by comparing the disease prediction performance against modality-specific datasets alone. To evaluate the impact of missing modalities in downstream tasks, instead, we compared prediction performance with and without MOFA+ imputation in the merged space. Our cohorts included only subjects with white ethnic background (i.e., the most frequent ethnic background in UKB data) to mitigate the potential effects of population stratification.\\
The baseline of our studies was set - within \textit{mmid} - to the first UKB imaging 
examination, 
when CMR and ECG tests were performed; subjects in the cohort without an available merged representation (i.e., without data for one or more modalities in the general case, or without data for any of the modalities in the case of MOFA+ fusion with imputation) were excluded from our disease prediction analyses by \textit{mmid}. Our package also automatically excluded from disease prediction the individuals having experienced endpoint-related events before baseline (\textbf{\autoref{tab:endpoint}}).\\
Given that AJIVE guarantees some degree of orthogonality between cross-modal dimensions (see\textbf{~\autoref{sec:fusion}}), we trained (1) penalised logistic regression models (with L1 regularisation) for disease classification, and (2) penalised Cox-PH models (with Elasticnet regularisation) for time-to-event prediction, aiming to forecast the future occurrence of CVD subtypes from the merged representation (in the multi-modal scenario) and from the first PCs - explaining $P=80\%$ of variance - of modality-specific datasets (for comparison). In the case of MOFA+ integration to deal with missing data sources, instead, we employed non-penalised logistic regression and Cox-PH models for prediction, because we decided to set to $10$ the maximum number of cross-modal features (see\textbf{~\autoref{sec:fusion}}). In both situations, 
the \textit{mmid} package
relied on the merged components as covariates of our incident disease prediction models because they provided a low-dimensional representation of the interactions between the considered modalities; when repeating the analyses with single-modality datasets, we adopted their PCs as features both to reduce their dimensionality and as a fairer comparison with the multi-modal scenario. 
Various combinations of input modalities, multi-modal fusion algorithms, and prediction methods and tasks were tested simply passing proper configuration files to the package's main function $mmid()$.

\subsection{\bf \it Model tuning and evaluation with mmid}
\label{sec:validation}
The disease prediction datasets were split within \textit{mmid} into 80\%-20\% train-test (i.e., $test\_size = 0.2$ as an argument to $mmid()$), preserving equal proportions of disease cases and controls in the two sets. Keeping the $20\%$ independent test aside, we employed $10$-fold cross-validation (i.e., $n\_folds = 10$) to select the best prediction model penalisation when necessary (by balancing between prediction performance and number of selected features), as well as to assess the added predictive value of multi-modal integration and/or missing modality imputation, always guaranteeing the original case-control ratios. We then tested the final models on our 20\% test set to further evaluate our results.

\section{\bf Application: Results}
\label{sec:RESULTS}
In this section, we present the results of our 
application study on the integration of cardiac imaging, ECG and genetic data for incident CVD prediction 
using our package. In particular, thanks to \textit{mmid} 
we could easily inspect how the three data sources contributed to the resulting cross-modal representation, 
and evaluate its prediction performance across CVD subtypes compared to the same predictions carried out using single modalities. 
We could 
finally assess 
the relevance of solving missingness of data sources in the latent integrated space, both in terms of increase in sample size and difference in performance. We focused our evaluations on the following three 
aspects: (1) the impacts of the analysed heterogeneous modalities on the cross-modal representation, (2) the value added to incident CVD prediction by each data source, and (3) the possibility to provide integration and prediction in the case of missing data and partially observable modalities. Again, all integration and prediction results presented in this section were obtained using the \textit{mmid} 
Python package.

\subsection{\bf \it AJIVE disentangles the joint and individual information captured by the modalities}
\label{sec:integration_results}
\begin{figure}[h]
\vspace{3mm}
 \begin{center}
 \includegraphics[width=0.65\textwidth]{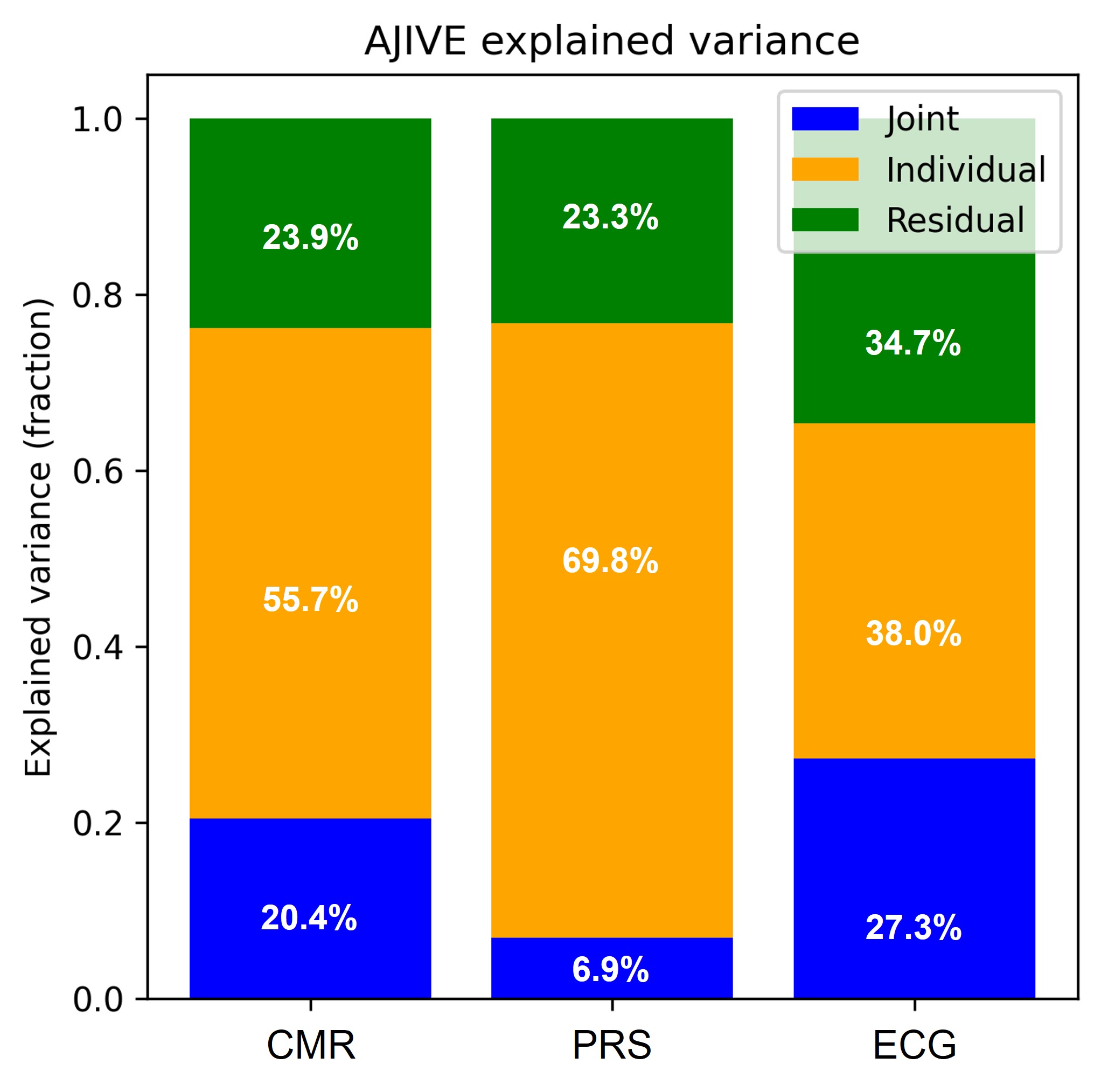}
\caption{\textbf{Variability explained by the AJIVE merged representation.} Variability explained by the joint and individual components (plus residual) of the Angle-based Joint and Individual Variation Explained merged representation, across the cardiac magnetic resonance (CMR) imaging, polygenic risk score (PRS) and electrocardiogram (ECG) datasets. This plot was created by the \textit{mmid} Python package.
\label{fig:ajive_integration}}
 \end{center}
\end{figure}

AJIVE projected the CMR, ECG and PRS datasets to a $49$-dimensional integrated space, available for 26,199 subjects having CMR, ECG, and PRS information recorded. Of the $M + \sum_{i=1}^{3} M_{i} = 49$ merged components, $M = 6$ captured joint effects shared by the three modalities, while $M_{1} = 20$, $M_{2} = 2$ and $M_{3} = 21$ were related to variability specific only to CMR-derived phenotypes, ECG features or genetic predisposition, respectively.\\
\textbf{\autoref{fig:ajive_integration}} shows that the CMR and ECG datasets shared the largest proportion of variance, while the effect of genetic features on the merged representation proved to be joint in a much more limited way. In any case, every modality showed to preserve its individual contribution and uniqueness, capturing effects not replaceable by the acquisition of the others. These individual contributions were dominant (i.e., capturing the largest proportion of variance) for all modalities, mostly for CMR and genetics.\\
Through the AJIVE weight matrices (see\textbf{~\autoref{sec:fusion}}), we were able to attribute the components of the merged representation to the original modality-specific features, as reported in\textbf{~\autoref{fig:ajive_interpretation_cmr}},\textbf{~\autoref{fig:ajive_interpretation_ecg}},\textbf{~\autoref{fig:ajive_interpretation_prs}}. As an example, we observed that the first joint AJIVE feature (\textit{Joint1}) was related, among others, to body size features (e.g., body surface area, aorta area, body mass index PRS); the first ECG individual component, instead, was mostly due to the PP interval and the number of QRS complexes during ECG acquisition.\\
In summary, we leveraged 
our \textit{mmid} package 
to exploit the interactions between CMR, ECG, and PRS datasets, as well as their intra- and inter-modality redundancies, and project the information into an interpretable merged representation. 
Thanks to the AJIVE method, 
we showed that the three modalities captured joint and individual information, which became the input for our disease prediction analyses.

\subsection{\bf \it AJIVE integration predicts the future occurrence of atrial arrhythmia}
\label{sec:ajive_prediction_results}
We predicted incident AA in a classification task and in a survival analysis framework, using the features of the AJIVE merged representation as covariates. 
We launched all analyses with our \textit{mmid} package, which performed the integration and prediction steps sequentially in a single run.\\
For classification, we considered a cohort of 12,159 subjects healthy at baseline, $467$ of whom ($3.8\%$) developed an AA event during a $5$-year follow-up. We compared the prediction performance of the merged representation of the three modalities against the PCs of single-modality datasets (i.e., CMR only, ECG only, PRS only).\textbf{~\autoref{tab:aa_results_table}} reports the corresponding prediction results, highlighting that the integration of the three modalities successfully predicted the future occurrence of AA in a healthy cohort in a $5$-year time horizon, performing significantly better than modality-specific datasets alone.\\
Specifically, we observed that the CMR modality was the one mostly impacting prediction performance, achieving $0.73$ cross-validation and test AUC when considered alone. In fact, $6$ individual CMR components of the AJIVE merged representation (out of $M_{1} = 20$) were also selected for AA prediction by our penalised logistic regression, with $3$ of them showing statistically significant effects in the model. Furthermore, logistic regression trained on the cross-modal representation selected 
$4$ out of $M = 6$ joint merged components, 
highlighting also the role of the shared contribution of the three modalities to disease risk. In addition to the relevance of CMR alone for incident AA prediction, we observed that its integration with information from an ECG test and genetic predisposition led to a significant improvement in prediction performance, according to a Wilcoxon signed rank test ($10\%$ level), i.e., $+1.5\%$ cross-validation AUC and $+2\%$ test AUC; fusing CMR features only with ECG ones or PRSs, instead, 
did not improve performance with statistical significance with respect to CMR alone, 
thus demonstrating that the integration of all three modalities together was key in this scenario.\\
Thanks to AJIVE interpretability and the penalised logistic regression model we employed, our predictions were fully explainable: as an example, the biggest effects on incident AA were due to two joint components of the cross-modal representation, one of which reducing the risk in the case of reduced myocardial wall thickness (from the CMR examination) and low genetic predisposition for high body mass index values and hypertension, and the other driven by the QRS duration measured in the ECG test and by the atrial fibrillation PRS (\textbf{\autoref{fig:ajive_interpretation_cmr}}, \textbf{\autoref{fig:ajive_interpretation_ecg}}, \textbf{\autoref{fig:ajive_interpretation_prs}}, the discussed features being \textit{Joint1} and \textit{Joint6}, respectively).\\
Our results were confirmed in the survival analysis framework, which aimed to predict time-to-AA in a population of 25,014 individuals healthy at baseline, with $565$ of them ($2.3\%$) developing AA during the follow-up (\textbf{\autoref{fig:kaplan_meier}(a)} shows the Kaplan-Meier plot for incident AA in the train set). In particular, the merged representation of the three modalities proved to be the best predictor of time-to-AA, outperforming the CMR dataset alone and the fusion of CMR and ECG: slightly in cross-validation (from $0.71$ to $0.72$ concordance index) and by nearly $+4\%$ concordance index in the test set (from $0.73$ to $0.77$).

\begin{table}[httb!] \small
\centering
    \begin{tabularx}{0.79\textwidth}{ c  c  c }
        \toprule
          \textbf{Modalities} & \textbf{Cross-validation AUC} & \textbf{Test AUC} \\
        \midrule
        \rowcolor{LightBlue} 
        \rowcolor{LightBlue} Cardiac magnetic resonance (CMR) \ding{61}\ding{73} & $0.73$ ($0.04$) & $0.73$ \\
        Electrocardiogram (ECG) \ding{73} & $0.62$ ($0.05$) & $0.59$ \\
        \rowcolor{LightBlue} Polygenic risk scores (PRS) & $0.57$ ($0.05$) & $0.61$ \\
        CMR + ECG + PRS *\ding{61}\ding{73} & $0.74$ ($0.05$) & $0.75$ \\
        \bottomrule
    \end{tabularx}
    \caption{\textbf{Atrial arrhythmia classification results.}
    Incident atrial arrhythmia classification results, in terms of area under the receiver operating characteristic curve (AUC), in cross-validation - mean (standard deviation) - and test using principal components of single modalities or their AJIVE merged features as covariates of a penalised logistic regression model. *, \ding{61} and \ding{73}, if present, indicate that the cross-validation AUC is statistically greater than that of cardiac magnetic resonance only, electrocardiogram only or polygenic risk scores only, respectively, according to a Wilcoxon signed rank test at $10\%$ level.\label{tab:aa_results_table}}
\end{table}

\subsection{\bf \it AJIVE representation is predictive across cardiovascular disease subtypes}
\label{sec:other_prediction_results}
We extended the classification and time-to-event prediction tasks to the following other cardiovascular disease subtypes: CAD, general CVD, and SHD. The results generally confirmed the relevance of the AJIVE merged representation in predicting the future occurrence of adverse events, highlighting the different roles of the modalities depending on the specific target disease.\\
For instance, in the case of CAD we observed a higher importance of PRSs over ECG compared to AA. Indeed, in the classification scenario (cohort of 11,970 subjects, with $381$ of them developing CAD in $5$ years), our penalised logistic regression model did not select any ECG individual features of the AJIVE representation, while $4$ joint (all significant), $11$ CMR individual and $7$ PRS individual components were selected with the L1 penaliser. Furthermore, genetic predisposition predicted incident CAD better than ECG in both classification ($0.61$ mean AUC in cross-validation and $0.65$ AUC in the test set by PRSs, versus $0.60$ and $0.61$ by ECG, respectively) and survival (\textbf{\autoref{tab:cad_results_table}} - in a cohort of 24,691 individuals with $430$ incident cases, Kaplan-Meier plot in\textbf{~\autoref{fig:kaplan_meier}(b)}), thus confirming our findings. Finally, analogously to the AA case, we achieved the best performance in classification and time-to-event prediction of future CAD events when considering the AJIVE merged representation of CMR, ECG and PRS information.\\
Regarding general CVD and SHD, instead, we already achieved robust prediction performance from the CMR dataset alone ($0.70$ mean AUC and $0.68$ mean concordance index in cross-validation for general CVD, while $0.76$ and $0.74$ for SHD, respectively), not (significantly) outperformed by leveraging the AJIVE integration, suggesting that cardiac imaging information was enough for prognostic prediction of these disease subtypes. In any case, integrating the three modalities with AJIVE led to non-detrimental classification and survival performance 
- both for general CVD ($0.71$ mean AUC and $0.68$ mean concordance index in cross-validation) and SHD ($0.76$ and $0.73$, respectively), 
thus still proving that multi-modal integration represented the three data sources in a shared space that could be successfully employed for disease prediction.

\begin{table}[httb!] \small
\centering
    \begin{tabularx}{0.85\textwidth}{ c  c  c }
        \toprule
          \textbf{Modalities} & \textbf{Cross-validation C-index} & \textbf{Test C-index} \\
        \midrule
        \rowcolor{LightBlue} 
        \rowcolor{LightBlue} Cardiac magnetic resonance (CMR) \ding{61}\ding{73} & $0.67$ ($0.04$) & $0.67$ \\
        Electrocardiogram (ECG) & $0.58$ ($0.03$) & $0.61$ \\
        \rowcolor{LightBlue} Polygenic risk scores (PRS) \ding{61} & $0.61$ ($0.06$) & $0.64$ \\
        CMR + ECG + PRS *\ding{61}\ding{73} & $0.69$ ($0.04$) & $0.70$ \\
        \bottomrule
    \end{tabularx}
    \caption{\textbf{Coronary artery disease time-to-event prediction results.}
    Incident coronary artery disease time-to-event prediction results, in terms of concordance index (C-index), in cross-validation - mean (standard deviation) - and test using principal components of single modalities or their AJIVE merged features as covariates of a penalised Cox proportional hazards model. *, \ding{61} and \ding{73}, if present, indicate that the cross-validation C-index is statistically greater than that of cardiac magnetic resonance only, electrocardiogram only or polygenic risk scores only, respectively, according to a Wilcoxon signed rank test at $10\%$ level.\label{tab:cad_results_table}}
\end{table}

\subsection{\bf \it MOFA+ solves missingness maintaining similar prediction performance}
\label{sec:missingness}
\begin{figure}[h]
\vspace{3mm}
 \begin{center}
 \includegraphics[width=0.95\textwidth]{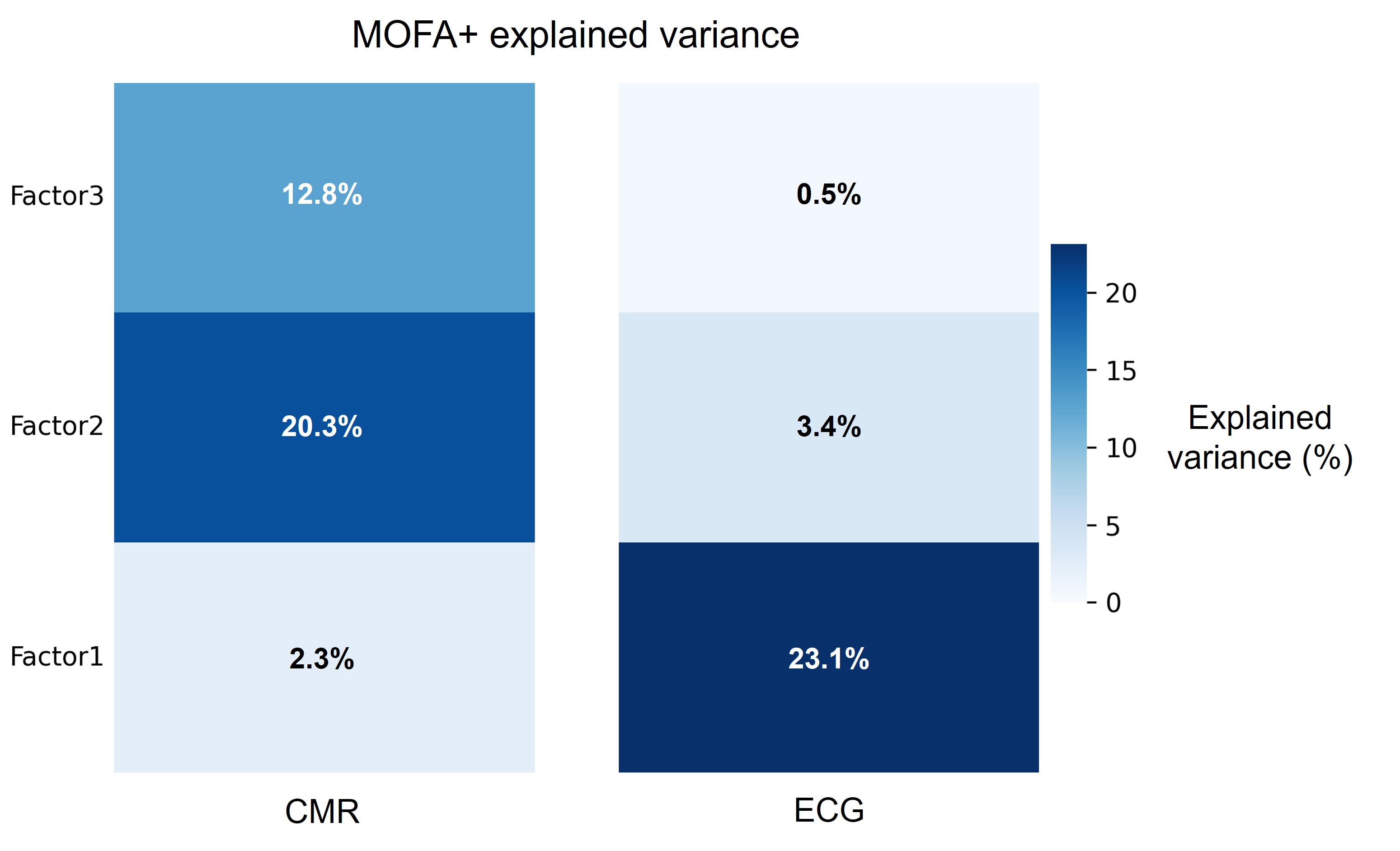}
\caption{\textbf{Variability explained by the MOFA+ merged representation.} Variability explained by the Multi-Omics Factor Analysis merged factors, across the cardiac magnetic resonance (CMR) imaging and electrocardiogram (ECG) datasets. The blue colour scale represents explained variance (\%). This plot was created by the \textit{mmid} Python package.
\label{fig:mofa_integration}}
 \end{center}
\end{figure}

To assess the integration and prediction performance in case of missing data modalities, we employed MOFA+ 
within the \textit{mmid} package 
to derive a cross-modal representation of CMR and ECG acquisitions to fit incident disease classification and survival analysis. Specifically, we investigated the relevance of modality imputation in the cross-modal space by MOFA+ integrating CMR and ECG data, thus not considering the PRS dataset, for the following main reasons: (1) CMR and ECG were the most correlated modalities (see\textbf{~\autoref{sec:integration_results}}), so imputation was expected to be more feasible compared to cases that included the PRS dataset (i.e., CMR+PRSs, ECG+PRSs, or CMR+ECG+PRSs); (2) contrary to CMR and ECG, genetic information was acquired for most UKB subjects, so considering PRSs would have implied imputing for the vast majority of the cohort while observing only a small fraction of individuals with all the modalities available. Taking into account only CMR and ECG data sources, 48,920 subjects had data for at least one modality available, and 26,906 of them had data for both (\textbf{\autoref{fig:mod_overlap}}).\\
MOFA+ projected the CMR and ECG datasets into a $3$-dimensional space (i.e., $M=3$), with the first dimension being mainly related to ECG information (but also capturing some CMR variability), and the other two mostly driven by CMR features (the third almost exclusively, while the second was partially shared with the ECG modality), as shown in\textbf{~\autoref{fig:mofa_integration}}. Analogously to AJIVE, also in this case we were able to inspect the factorisation weight matrices to interpret the relationships between the cross-modal dimensions and the original features of the CMR and ECG datasets (\textbf{\autoref{fig:mofa_interpretation}}).\\
The \textit{mmid} package 
then used the MOFA+ $3$-dimensional merged representation to fit disease classification and time-to-event prediction analyses across AA, CAD, general CVD and SHD. Given the low dimensionality of the input, we decided not to penalise our logistic regression and Cox-PH models. We compared the cases without imputation of missing modalities (i.e., cohorts of only individuals with both CMR and ECG acquired) versus imputation (i.e., all subjects with either CMR or ECG, or both, collected).\textbf{~\autoref{tab:missingness_results_table}} shows that, when imputing missing modalities in the integrated space, we were able to extend our predictions to significantly larger cohorts (i.e., from about 12k to 18k subjects for classification, and from nearly 25k to 45k in time-to-event analyses), without considerable losses in performance. We thus demonstrated that MOFA+ imputation 
within \textit{mmid} 
allowed us not only to predict even with partially acquired modality data, but to do it as if both modalities were observed for all individuals.\\
Again, we used MOFA+ and not AJIVE in this situation because the latter was not designed to deal with missing modalities. In all previous case study analyses that did not involve imputation, instead, we preferred AJIVE due to its easier interpretation, explicitly disentangling integrated information into joint and modality-specific components, and generally superior performance in prediction tasks.

\begin{table}[httb!] \small
\centering
    \begin{tabularx}{1.04\textwidth}{ c || c  c | c  c }
        \toprule
          \textbf{\CellWithForcedBreak{Disease \\ subtype}} & \textbf{\CellWithForcedBreak{[clf no imputation] \\ AUC CV \\ AUC test \\ Cohort size}} & \textbf{\CellWithForcedBreak{[clf imputation] \\ AUC CV \\ AUC test \\ Cohort size}} & \textit{\textbf{\CellWithForcedBreak{[tte no imputation] \\ C-index CV \\ C-index test \\ Cohort size}}} & \textit{\textbf{\CellWithForcedBreak{[tte imputation] \\ C-index CV \\ C-index test \\ Cohort size}}} \\
        \midrule
        \midrule
        \rowcolor{LightBlue} 
        \rowcolor{LightBlue} \CellWithForcedBreak{Atrial \\ arrhythmia} & \CellWithForcedBreak{$0.65$ ($0.07$) \\ $0.66$ \\ 12,159} & \CellWithForcedBreak{$0.64$ ($0.03$) \\ $0.63$ \\ 18,793} & \CellWithForcedBreak{$0.65$ ($0.03$) \\ $0.65$ \\ 25,014} & \CellWithForcedBreak{$0.65$ ($0.04$) \\ $0.62$ \\ 45,032} \\
        \CellWithForcedBreak{Coronary \\ artery \\ disease} & \CellWithForcedBreak{$0.69$ ($0.05$) \\ $0.69$ \\ 11,970} & \CellWithForcedBreak{$0.68$ ($0.04$) \\ $0.67$ \\ 18,435} & \CellWithForcedBreak{$0.69$ ($0.04$) \\ $0.70$ \\ 24,691} & \CellWithForcedBreak{$0.67$ ($0.03$) \\ $0.70$ \\ 44,593} \\
        \rowcolor{LightBlue} \CellWithForcedBreak{Cardiovascular \\ disease} & \CellWithForcedBreak{$0.65$ ($0.03$) \\ $0.65$ \\ 11,916} & \CellWithForcedBreak{$0.63$ ($0.02$) \\ $0.65$ \\ 18,411} & \CellWithForcedBreak{$0.65$ ($0.03$) \\ $0.61$ \\ 23,957} & \CellWithForcedBreak{$0.64$ ($0.01$) \\ $0.60$ \\ 43,092} \\
        \CellWithForcedBreak{Structural \\ heart \\ disease} & \CellWithForcedBreak{$0.70$ ($0.06$) \\ $0.65$ \\ 12,162} & \CellWithForcedBreak{$0.70$ ($0.04$) \\ $0.65$ \\ 18,805} & \CellWithForcedBreak{$0.70$ ($0.05$) \\ $0.66$ \\ 25,188} & \CellWithForcedBreak{$0.68$ ($0.03$) \\ $0.71$ \\ 45,562} \\
        \bottomrule
    \end{tabularx}
    \caption{\textbf{MOFA+ dealing with missing modalities: prediction results from cardiac magnetic resonance imaging and electrocardiogram data integration.}
    Incident disease classification (clf) and time-to-event (tte) prediction results from cardiac magnetic resonance imaging and electrocardiogram MOFA+ integration with and without imputation of missing modalities in the cross-modal space, in terms of area under the receiver operating characteristic curve (AUC) and concordance index (C-index), in cross-validation (CV) - mean (standard deviation) - and test. MOFA+ merged factors were used as covariates of logistic regression and Cox proportional hazards models.\label{tab:missingness_results_table}}
\end{table}

\section{\bf Discussion}
\label{sec:discussion}
This works shows the relevance of multi-modal fusion for integrating multiple health data sources in a shared representation, which can be used in particular for 
enhancing more precise risk stratification and improving both classification tasks and survival predictions. For this purpose, we propose 
\textit{mmid}, a 
Python package that allows integrating 
tabular datasets and 
taking advantage of 
the corresponding cross-modal representation for a classification, survival or clustering task. In particular, the possibility to sequentially perform data integration and a downstream analysis 
in a unique package and through a single function call, with a selection of state-of-the-art algorithms for both the steps, easily allows evaluating the impact of multi-modal fusion and data imputation on the task, as well as comparing different methods and robustifying analyses of complex multi-modal healthcare data. This is key especially in the framework of integration, which is intrinsically hard to evaluate per se, but gains relevance as a data preprocessing/preparation step before the downstream task of interest. We believe that \textit{mmid} bridges the gap between standalone multi-modal integration algorithms (and corresponding libraries) and the need to compare their outputs and use them quickly and easily for a variety of disease prediction and/or patient clustering analyses. Indeed, the 
package enables access to state-of-the-art fusion methods in a unified way, guaranteeing transferability, easier reproducibility, and efficiency of two-step (i.e., integration plus downstream) multi-modal analyses. 
Furthermore, \textit{mmid} is extensible, allowing novel data fusion and downstream algorithms to be easily included in its future versions.\\
We showed the relevance of \textit{mmid} through an 
example usage 
application about the integration of cardiac imaging, ECG and genetic data for incident CVD prediction. We chose this case study as it exemplifies a context in which our 
package could be used to tackle the need for efficient integration of diverse health data sources and disease prediction as if it were a single-step analysis, even in the case of missing views. Furthermore, the application scenario required the evaluation of the added value of single modalities: to do so, we heavily relied on our 
package to standardise analytical procedures and replicate the analyses using single modalities alone, couples of them combined, or the integration of all the three.\\
First, through \textit{mmid}, we were able to explicitly represent and interpret the relationships between cardiac imaging measures, ECG features, and genetic predisposition, showing that the modalities capture both individual and partially overlapping information that can be particularly beneficial for disease prediction when integrated together. 
Thanks to the possibility to easily replicate analyses with different input data modalities, we used \textit{mmid} to assess the value of the three analysed data sources and their integration in our downstream tasks, across endpoints. 
In particular, in the predictions of incident cardiovascular events in subjects healthy at baseline from the merged representation obtained with the AJIVE methodology, we 
validated the 
general relevance of merged components in the models
. In addition to the latter, also 
modality-specific contributions were shown to play a role in the downstream task, 
clarifying that individual information captured by CMR alone, ECG alone, or genetics alone could not be replaced a priori by the other tests. 
Our results highlighted that integrating multiple health data sources for disease prediction is often beneficial, and anyway never detrimental, compared to analysing modalities alone or subsets of them: 
we observed that multi-modal integration contributed to early identification of future patients, with clinically valuable performance (i.e., always close or higher than $0.70$ AUC and concordance index), thus opening the doors for personalised CVD prognosis before the appearance of symptoms and clinical manifestations.\\
With the rising availability of big healthcare data from different examinations in clinical practice, we believe that multi-modal integration might play an increasingly relevant role in the efficient representation of such large-scale complex information, and in its employment for different tasks in various biological and medical domains. In such a context, in which large populations may undergo different examinations depending on their characteristics and health status, the possibility of integrating data modalities and performing downstream tasks even in the absence of some medical tests would possibly extend the impacts of multi-modal fusion to larger cohorts; in our work, we showed that it is possible to impute missing data modalities in the cross-modal space with \textit{mmid} without considerable performance losses, thus 
validating the use of our framework in real-case scenarios.\\
A limitation of 
\textit{mmid} - and, in turn, of our application study - is the dependence of the merged representation on the type and number of features extracted by the raw modalities. Indeed, this may 
overestimate or underestimate the importance of some modalities in the data fusion and disease prediction steps. 
We mitigated this effect by reducing the collinearity of the modality-specific input dataset (i.e., iteratively removing features with a variance inflation factor $> 10$, see\textbf{~\autoref{sec:data}}) and by reducing their dimensionality (i.e., cutoffs for explained variance, see\textbf{~\autoref{sec:integration}}); however, despite such steps are recommended, they may not prevent underestimations/overestimations due to redundant modalities and/or different variability scales of single data sources. 
Another limiting factor is related to the fact that 
our package accepts only tabular datasets as inputs. Finally, if input data come from different sources, they may require some harmonisation procedures (e.g., removal of batch effect and variability for imaging acquisitions from different scanners/protocols/sites~\cite{caldera2025scanneragnosticmriharmonizationssimguided}) which go beyond the scopes of \textit{mmid}.\\
In this work, \textit{mmid} was tested on an application scenario that highlighted the potential of integrating CMR, ECG and PRS data for cardiovascular risk prediction. Future work will try to further validate our findings with different types of data modalities and endpoints, possibly showing that (1) the approach can be generalised and potentially adopted in multiple clinical situations (e.g., other downstream analysis types, diseases, data modalities, and populations), and (2) \textit{mmid} allows again to smoothly tackle the tasks in a single environment.

\section{\bf Conclusion}
\label{sec:CONCLUSIONS}
In this work, we presented \textit{mmid}, a Python package offering a unified interface for multi-modal data fusion and imputation in healthcare, with potential applications to disease classification, clustering and time-to-event prediction.
We demonstrated the relevance of the package and of multi-modal data integration to combine heterogeneous data sources for downstream tasks, with an 
example 
application about the prediction of the future occurrence of cardiovascular events in healthy individuals. Indeed, we showed that it is possible to project modality-specific datasets into a cross-modal representation through data fusion methods, thus explicitly leveraging the interactions between different types of data coming from medical tests towards personalised risk profiling (and potentially other relevant downstream tasks). Furthermore, our results proved that the integration of CMR, ECG, and genetic predispositions is profitable in both $5$-year incident disease classification and time-to-event prediction across four tested CVD subtypes. Finally, we believe that the discussed 
possibility to deal with missing modalities with specific data fusion algorithms (e.g., MOFA+) might enhance the clinical translation of the proposed framework when dealing with complex high-dimensional health data sources across various downstream tasks. In a real clinical scenario in which new CMR, ECG, and PRS data points from different (i.e., testing) cohorts are available, the trained multi-modal integration models should be applied on these testing data to obtain their corresponding cross-modal representations; the latter would then be passed to the trained prediction models, which would output the risks associated to the testing dataset. Clearly, the effectiveness of the trained models is expected to be lower if the training and testing datasets come from different data distributions.\\
All analyses presented in this work were performed with the \textit{mmid} 
Python package, which sequentially employs state-of-the-art multi-modal fusion algorithms to factorise the diverse input datasets into a fully interpretable integrated low-dimensional space, followed by disease prediction (or patient clustering) from the latter. Despite having been applied to specific 
real 
scenarios 
(i.e., CVD prognosis with CMR, ECG, and PRS data integration), the 
package is general and can be used in any health analytics task involving the use of diverse high-dimensional tabular data sources and a prediction or clustering analysis on a selected cohort. Indeed, 
the results achieved in our case study may be successfully replicated when dealing with other 
health data sources and different diseases. Our 
package stands as a first step towards the standardisation of the use of multi-view integration algorithms as a 
necessary analysis preceding relevant downstream tasks, promoting more reproducible and transferable studies and results in the multi-modal healthcare analytics domain. Eventually, we believe that employing multi-modal integration to combine the increasingly available and 
heterogeneous 
health data sources collected nowadays could enhance better informed analytics and personalisation for a variety of downstream clinical and biological problems and tasks.

\section*{\bf Conflict of interests}
\label{sec:CONFLICT-OF-INTERESTS}
The authors declare that they have no known competing financial interests or personal relationships that could have 
influenced the work reported in this paper.

\section*{\bf Acknowledgments}
\label{sec:ACKNOWLEDGMENTS}
This research has been conducted using the UK Biobank Resource under application number 82779.\par
AMV, VI, FI acknowledge the MUR Excellence Department Project 2023-2027 awarded to the Department of Mathematics, Politecnico di Milano.\par
The authors would like to thank Carlo Andrea Pivato for his support in the definition of atrial arrhythmia and structural heart disease based on hospital diagnoses and causes of death.

\section*{Authors contributions}
AMV developed the \textit{mmid} Python package and performed the analyses of the application study.\\
VI supported the development of the \textit{Multi-modal fusion} module of \textit{mmid} and some application analyses.\\
EDA, MM and FI supervised the study.\\
AMV wrote and edited the manuscript. All authors read and reviewed the manuscript.

\section*{\bf Code availability}
\label{sec:AVAILABILITY}
\textit{mmid} is available at \url{https://github.com/ht-diva/mmid}

\footnotesize
\bibliographystyle{unsrt} 
\bibliography{bibliography.bib}
\normalsize

\clearpage
\section*{\bf Supplementary tables}
\label{sec:sup_t}
\makeatletter
\renewcommand \thetable{S\@arabic\c@table}
\makeatother

\begin{table}[httb!] \small
\centering
    \begin{tabularx}{0.95\textwidth}{ c || c | c }
        \toprule
          \textbf{\CellWithForcedBreak{Disease \\ subtype}} & \textbf{\CellWithForcedBreak{Defined as the first \\ among these events}} & \textbf{\CellWithForcedBreak{Excluding subjects that experienced \\ any of these events before baseline}} \\
        \midrule
        \midrule
        \\ AA & \textbf{ICD-10:} I48 & \textbf{ICD-10:} I48 \\ \\
        \hline 
        \\ CAD & \CellWithForcedBreak{\textbf{ICD-10:} I21, I22, I23, I24.1, I25.2 \\ \\ \textbf{ICD-9:} 410, 411, 412, 429.79 \\ \\ \textbf{OPCS-4:} K40.1-4, K41.1-4, K45.1-5, \\ K49.1-2, K49.8-9, K50.2, \\ K75.1-4, K75.8-9} & \CellWithForcedBreak{\textbf{ICD-10:} I21, I22, I23, I24.1, I25.2 \\ \\ \textbf{ICD-9:} 410, 411, 412, 429.79 \\ \\ \textbf{OPCS-4:} K40.1-4, K41.1-4, K45.1-5, \\ K49.1-2, K49.8-9, K50.2, \\ K75.1-4, K75.8-9} \\ \\
        \hline
        \\ CVD & \CellWithForcedBreak{\textbf{ICD-10:} I20, I21, I22, I23, \\ I24, I25, I50, I60, I61, I62, I63, \\ I64, I65, I66, I67, I68, I69 \\ \\ \textbf{ICD-9:} 410, 411, 412, 413, 414, 428, \\ 430, 431, 432, 433, 434, 436, 437, 438} & \CellWithForcedBreak{\textbf{ICD-10:} F01, I20, I21, I22, I23, \\ I24, I25, I50, I60, I61, I62, I63, \\ I64, I65, I66, I67, I68, I69 \\ \\ \textbf{ICD-9:} 410, 411, 412, 413, 414, 428, \\ 430, 431, 432, 433, 434, 436, 437, 438} \\ \\
        \hline
        \\ SHD & \CellWithForcedBreak{\textbf{ICD-10:} I42.0, I42.1, I42.2, I42.8, \\ I42.9, I46, I49.0, I50, I51.7}  & \CellWithForcedBreak{\textbf{ICD-10:} I42.0, I42.1, I42.2, I42.8, \\ I42.9, I46, I49.0, I50, I51.7} \\ \\
        \bottomrule
    \end{tabularx}
    \caption{\textbf{Endpoints definition.}
    Definition of incident atrial arrhythmia (AA), coronary artery disease (CAD), cardiovascular disease (CVD), and structural heart disease (SHD) cohorts in the UK Biobank study based on ICD-10 (for hospital diagnoses and causes of death), ICD-9 (for hospital diagnoses) and OPCS-4 codes (for operative procedures).\label{tab:endpoint}}
\end{table}

\clearpage
\section*{\bf Supplementary figures}
\label{sec:sup_f}
\makeatletter
\renewcommand \thefigure{S\@arabic\c@figure}
\makeatother

\begin{figure}[h]
\vspace{3mm}
 \begin{center}
 \includegraphics[width=\textwidth]{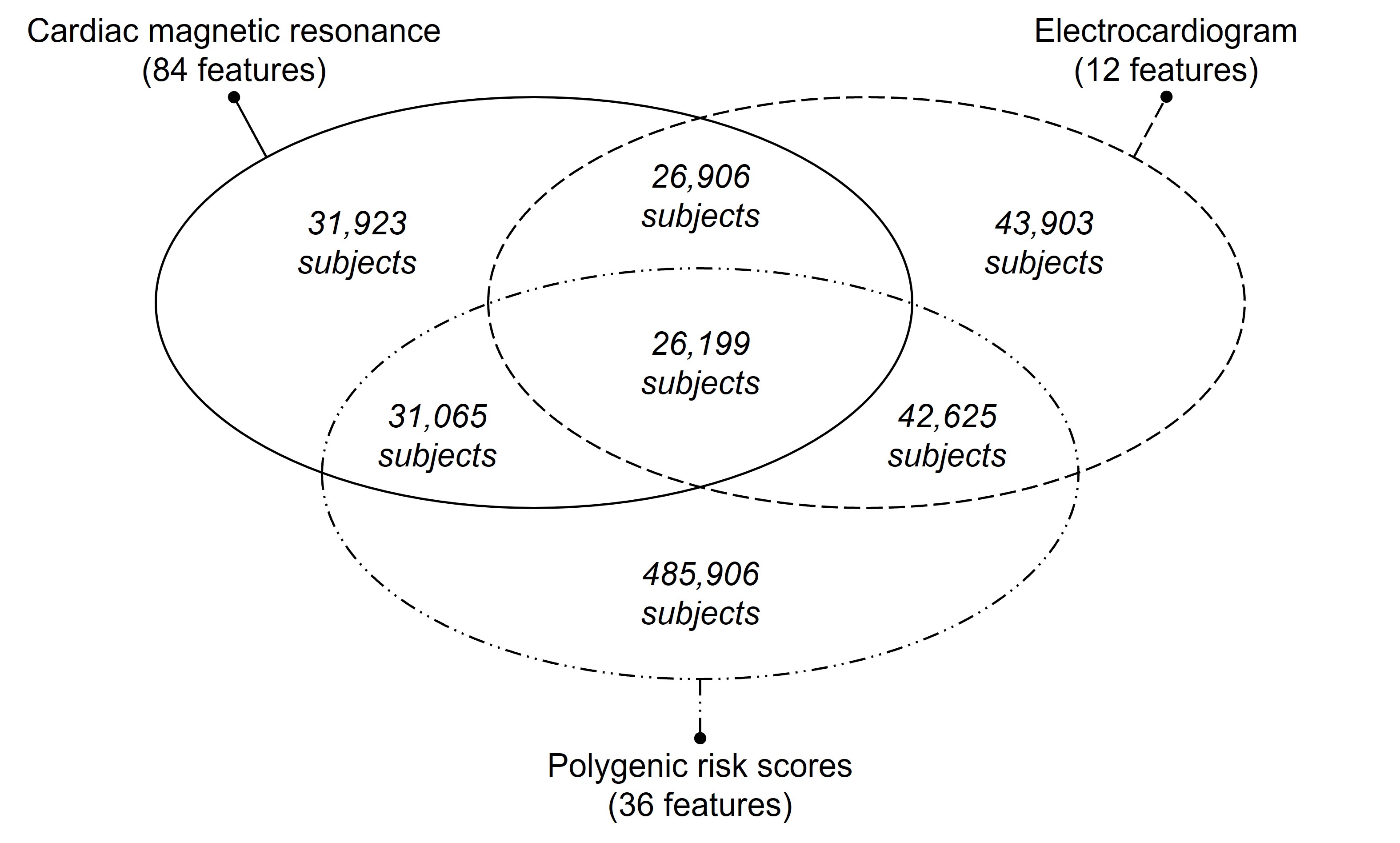}
\caption{\textbf{Details on the analysed modalities.} Number of features and subjects per data modality, including overlap information.
\label{fig:mod_overlap}}
 \end{center}
\vspace{-8mm}
\end{figure}


\begin{figure}[h]
\vspace{3mm}
 \begin{center}
 \includegraphics[width=0.55\textwidth]{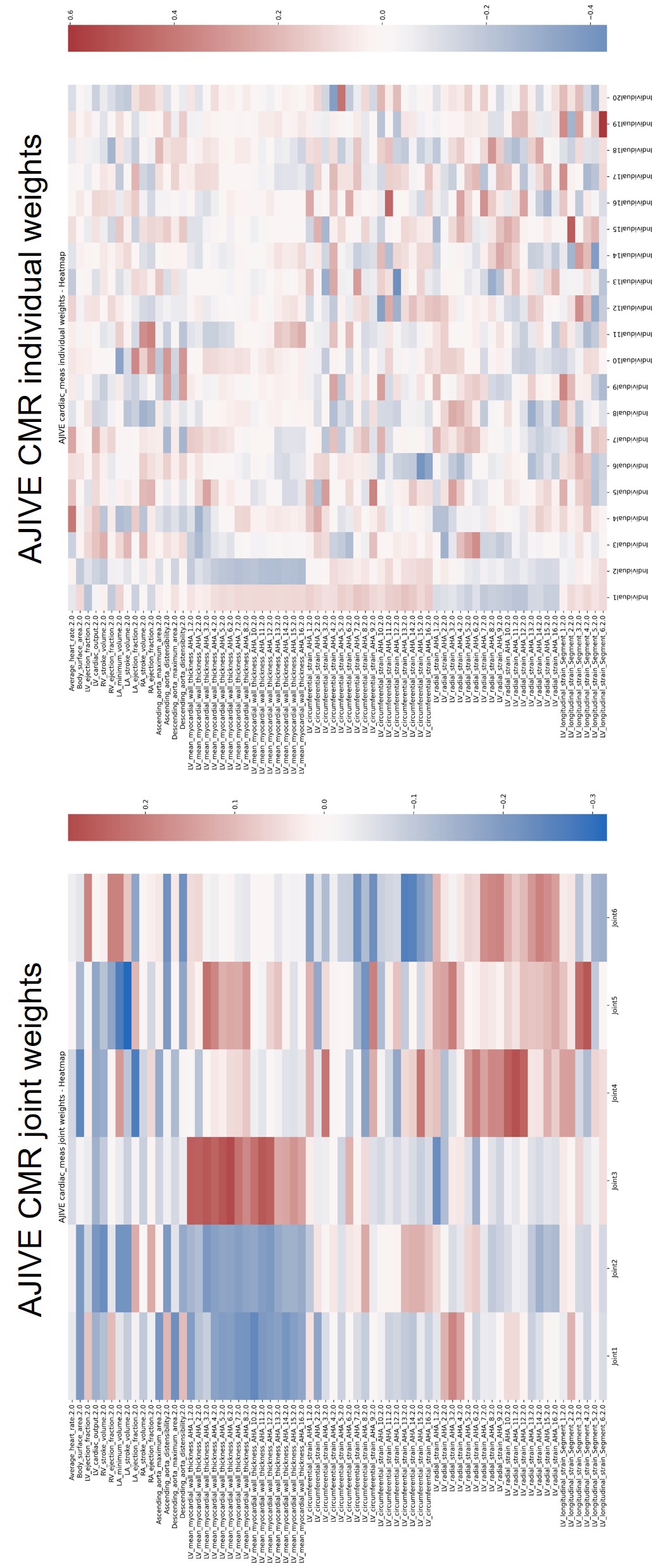}
\caption{\textbf{AJIVE interpretation heatmaps for the cardiac magnetic resonance (CMR) imaging dataset.} AJIVE interpretation heatmaps relating the joint and individual components of the cross-modal representation to the features of the input dataset. The colour scale represents the value of the weights in the integration weight matrices. These plots were created by the \textit{mmid} Python package.
\label{fig:ajive_interpretation_cmr}}
 \end{center}
\vspace{-8mm}
\end{figure}

\begin{figure}[h]
\vspace{3mm}
 \begin{center}
 \includegraphics[width=0.55\textwidth]{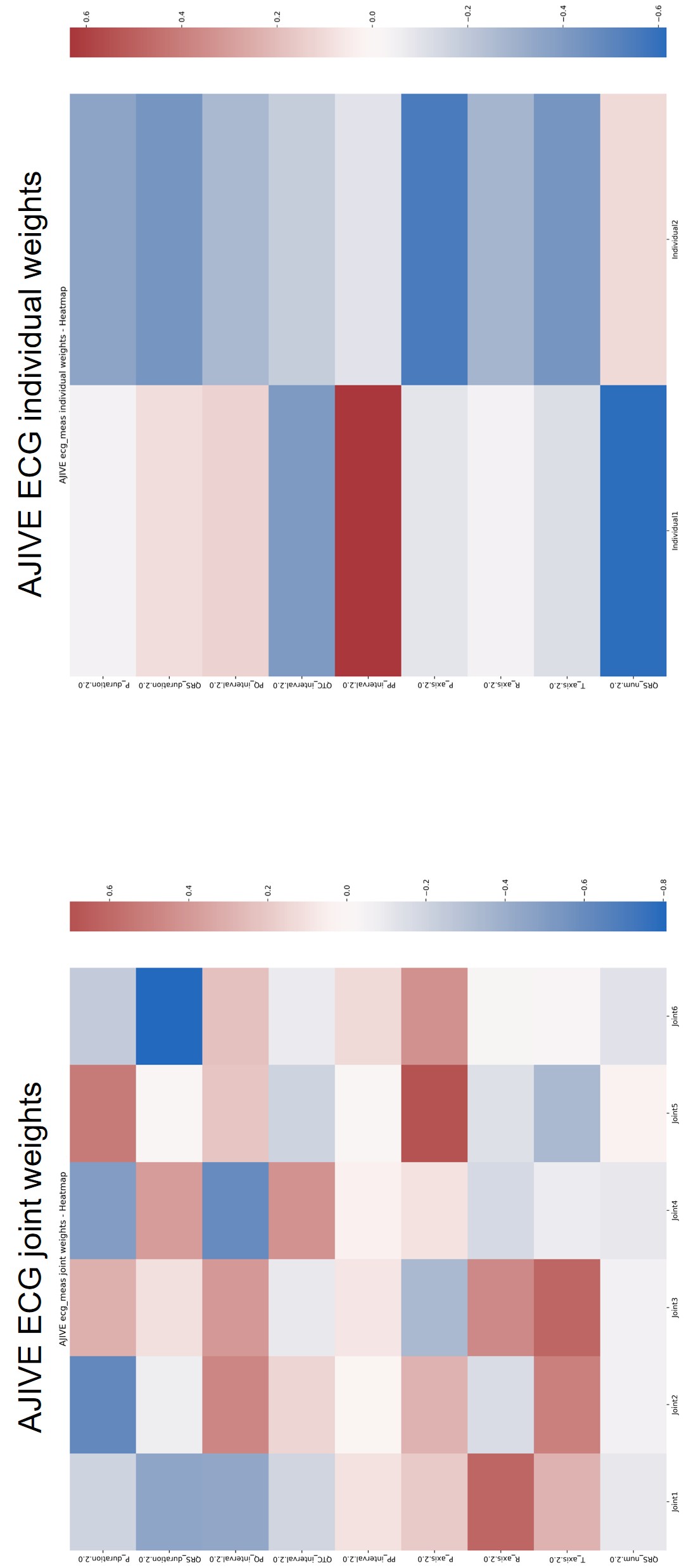}
\caption{\textbf{AJIVE interpretation heatmaps for the electrocardiogram (ECG) dataset.} AJIVE interpretation heatmaps relating the joint and individual components of the cross-modal representation to the features of the input dataset. The colour scale represents the value of the weights in the integration weight matrices. These plots were created by the \textit{mmid} Python package.
\label{fig:ajive_interpretation_ecg}}
 \end{center}
\vspace{-8mm}
\end{figure}

\begin{figure}[h]
\vspace{3mm}
 \begin{center}
 \includegraphics[width=0.55\textwidth]{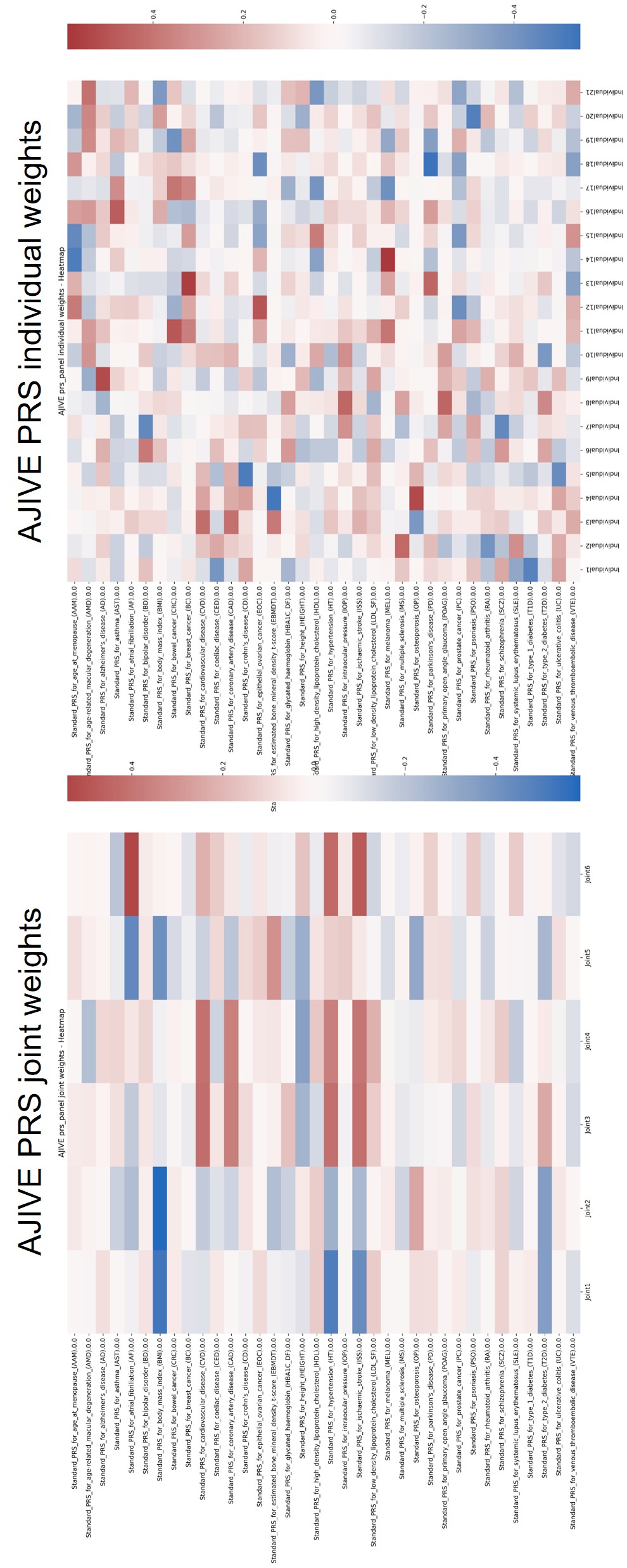}
\caption{\textbf{AJIVE interpretation heatmaps for the polygenic risk score (PRS) dataset.} AJIVE interpretation heatmaps relating the joint and individual components of the cross-modal representation to the features of the input dataset. The colour scale represents the value of the weights in the integration weight matrices. These plots were created by the \textit{mmid} Python package.
\label{fig:ajive_interpretation_prs}}
 \end{center}
\vspace{-8mm}
\end{figure}

\begin{figure}[h]
\vspace{3mm}
 \begin{center}
 \includegraphics[width=0.8\textwidth]{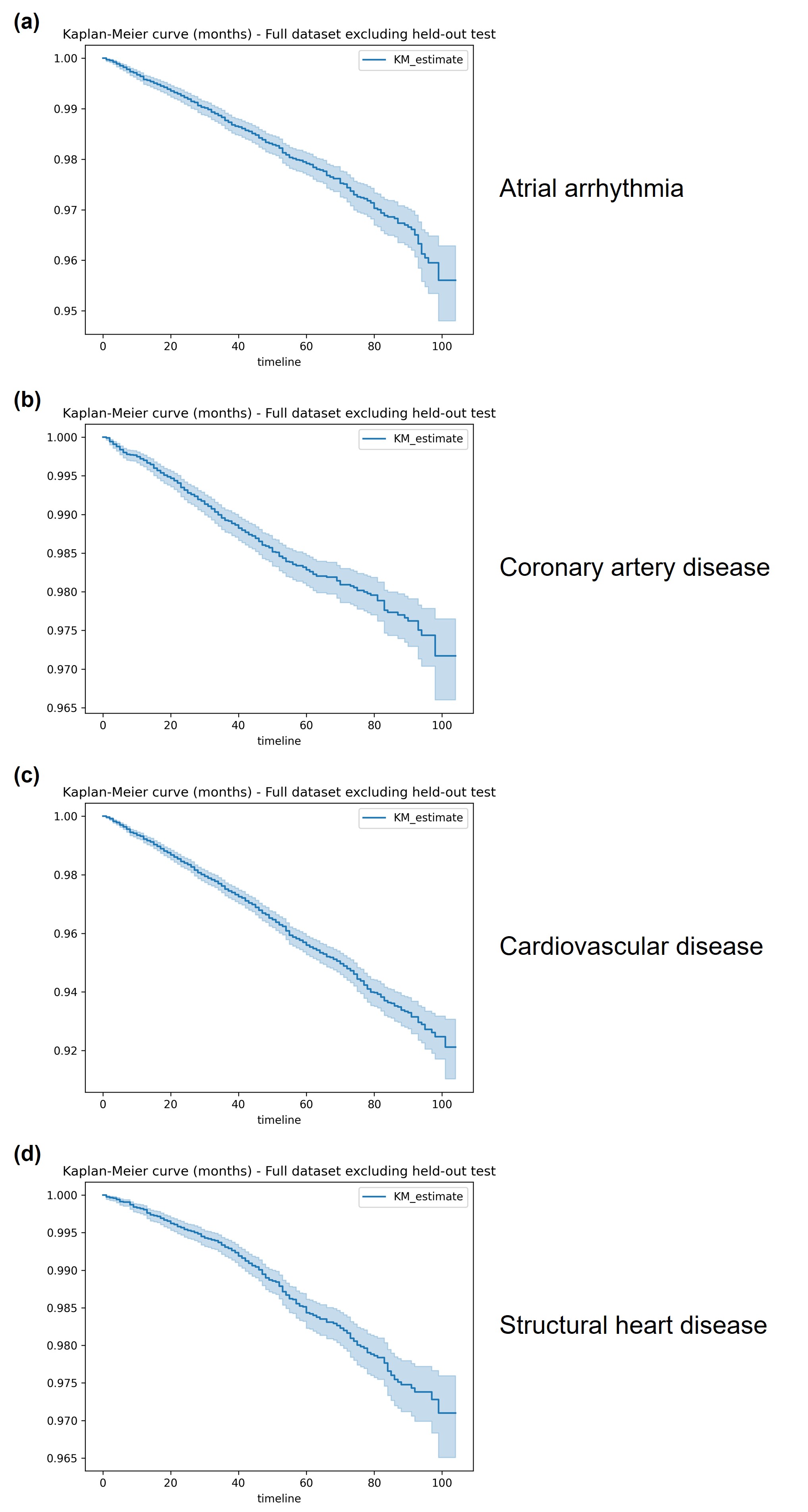}
\caption{\textbf{Kaplan-Meier (KM) survival curves across cardiovascular disease subtypes.} (a) Atrial arrhythmia. (b) Coronary artery disease. (c) General cardiovascular disease. (d) Structural heart disease.
\label{fig:kaplan_meier}}
 \end{center}
\vspace{-8mm}
\end{figure}

\begin{figure}[h]
\vspace{3mm}
 \begin{center}
 \includegraphics[width=\textwidth]{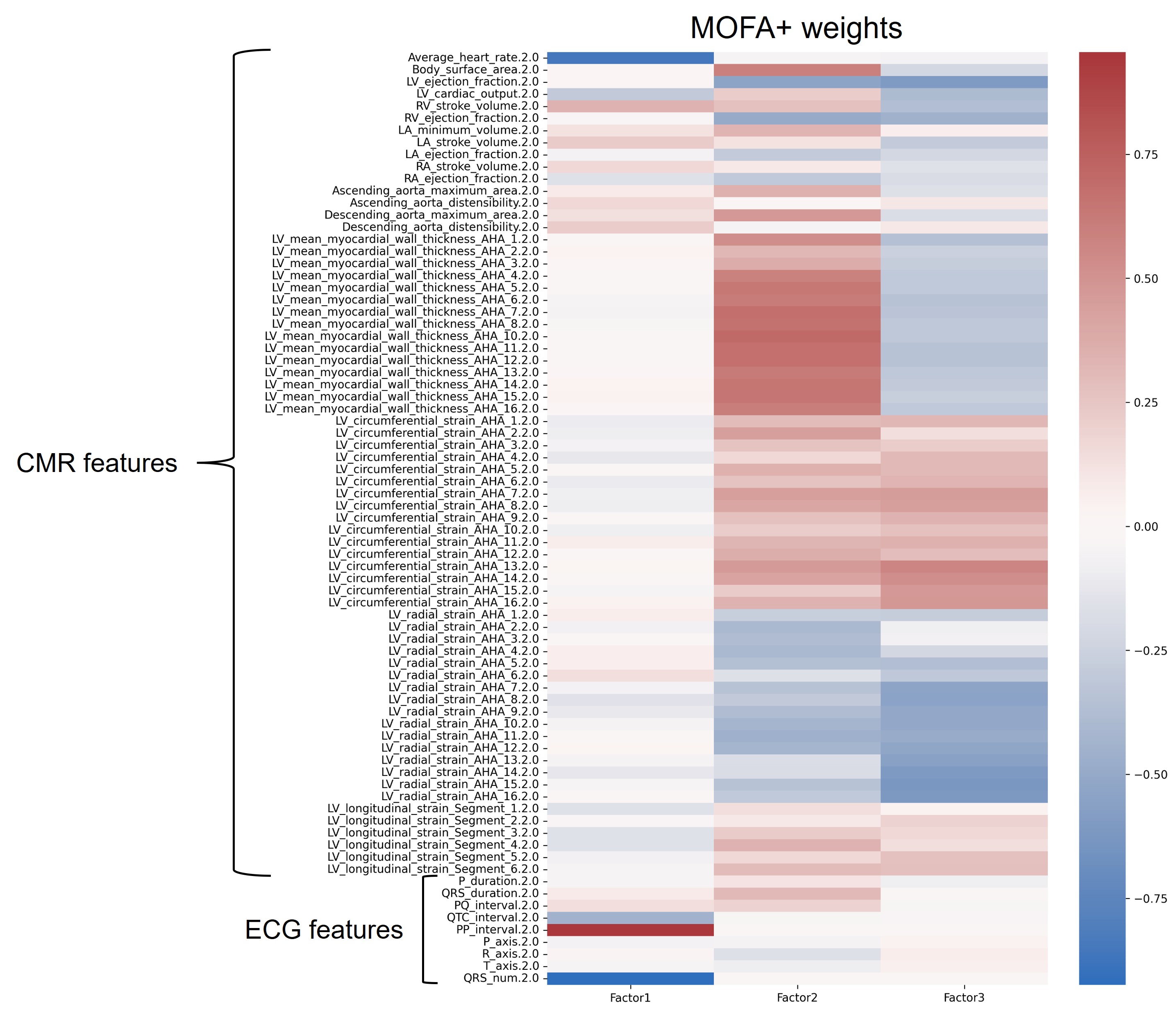}
\caption{\textbf{MOFA+ interpretation heatmap based on the integration weight matrices.} MOFA+ interpretation heatmap relating the factors of the cross-modal representation to the features of the input cardiac magnetic resonance (CMR) imaging and electrocardiogram (ECG) datasets. The colour scale represents the value of the weights in the integration weight matrices. This plot was created by the \textit{mmid} Python package.
\label{fig:mofa_interpretation}}
 \end{center}
\vspace{-8mm}
\end{figure}

\end{document}